\documentstyle[12pt,epsf]{article}
\newcommand{\sect}[1]{\section{#1}\setcounter{equation}{0}}

 \newcommand{\r}{\rightarrow}
 \newcommand{\be}{\begin{equation}}
 \newcommand{\ee}{\end{equation}}
 \newcommand{\eel}[1]{\label{#1}\end{equation}}
 \newcommand{\bea}{\begin{eqnarray}}
 \newcommand{\eea}{\end{eqnarray}}
 \newcommand{\eeal}[1]{\label{#1}\end{eqnarray}}
 \newcommand{\baq}{\begin{equation}\begin{array}{rcl}}
 \newcommand{\eaq}{\end{array}\end{equation}}
  \newcommand{\eaql}[1]{\end{array}\label{#1}\end{equation}}
  \newcommand{\beac}{\begin{equation}\begin{array}{rcl}}
  \newcommand{\eeacn}[1]{\end{array}\label{#1}\end{equation}}
  \newcommand{\ba}{\begin{array}}
  \newcommand{\ea}{\end{array}}
  \newcommand{\non}{\nonumber \\}
  
  \renewcommand{\a}{\alpha}
  \renewcommand{\b}{\beta}
             
           \newcommand{\D}{\Delta}

  \newcommand{\al}{{\alpha'}}
  \newcommand{\beq}{\begin{eqnarray}}
  \newcommand{\eeq}{\end{eqnarray}}

  \newcommand{\w}{Schwarzschild $\:$}

\begin{document}
  \begin{titlepage}
  
  \begin{flushright}
  hep-th/9901012
  \end{flushright}
  \vspace{2cm}
  
  \begin{center}
 \LARGE{\bf Black Holes, Shock Waves, and Causality
\\ in the AdS$/$CFT Correspondence}

  \vspace{10mm}
  
  \normalsize{Gary T. Horowitz \footnote{ gary@cosmic.physics.ucsb.edu}
 and  N. Itzhaki \footnote{ sunny@solkar.physics.ucsb.edu}}\\
  
  \vspace{.7cm}
  \normalsize{
  Department of Physics\\
  University of California, Santa Barbara, CA 93106}

  \end{center}
  \vskip 0.61 cm
  
\begin{abstract}

We find the expectation value of the energy-momentum tensor in the 
CFT corresponding to a moving black hole
in AdS. Boosting the black hole to the speed of light, keeping the total
energy fixed,
yields a gravitational shock wave  in AdS. The analogous procedure
on the field theory side leads to ``light cone'' states, i.e., states
with energy-momentum tensor localized on the light cone.
The correspondence between the gravitational shock wave and these light cone
states provides a useful tool for testing causality. 
 We show, in several examples, how the
CFT reproduces the causal relations in AdS.

\end{abstract}
\end{titlepage} 

\newpage

\baselineskip 20pt
\sect{Introduction}

Most of the recent investigations of the correspondence between
string theory in anti-de Sitter (AdS) space and conformal field theory
(CFT) \cite{juan} have focused on the Euclidean regime. Lorentzian processes
have just begun to be studied \cite{law,bdhm}.
We will be concerned with the question of how some basic causal 
relations in AdS are reproduced
 in the conformal field theory.
In the process of answering this question, we are led to a new description
of black holes in terms of the CFT, and an interesting
connection between gravitational shock waves in AdS and novel states
of the CFT which are localized on the light cone.

As an example of the type of situation we wish to analyze,
consider two massless particles which come from infinity in AdS
from the same direction,  but at different times. It is clear that
the later particle cannot influence the earlier one since it is entirely 
to its future. However, in the CFT, 
the ``scale-radius duality" \cite{sus}
suggests that a single massless particle that
comes in from infinity should be described by a localized excitation
that expands outward \cite{bdhm}. So  two massless particles should be
described by two excitations, but since the particles start at the same
position at different times, the second excitation lies inside
the light cone of the first. It is certainly not obvious why
the presence of the second excitation cannot influence the further
evolution of the first.

As a second example, suppose we consider two  
massless particles that come in at the same time, but from different directions.
These particles
should be described by two localized excitations in the CFT which
start at different points in space at the same time. If the excitations
are separated by a distance $c$, it is
clear from the CFT that interactions between them cannot occur before
a time $t = c/2$, and are very likely thereafter.
Is there an analogous statement on the AdS side? At first sight this seems
unlikely since the two massless particles need not intersect, and
even if they do, it is typically at a much later time.

We will resolve these puzzles below, and show that there is perfect
agreement between the answers one obtains in the CFT and AdS.
The first step is to realize that one
must take into account the gravitational backreaction of the massless
particles. For zero cosmological constant, the gravitational backreaction
is given by the Aichelburg-Sexl metric \cite{sexl} and
describes a gravitational shock wave. This solution can be obtained
by boosting the Schwarzschild metric with mass $M$ and taking the limit
as $M\rightarrow 0, \ v\rightarrow 1$ keeping the total energy fixed.
For negative cosmological constant, one can do exactly the same thing
starting with the Schwarzschild anti-de Sitter metric. This was first
done in four dimensions in \cite{swads} and is generalized to $d$ dimensions
below (see also \cite{new2}).
  It is known that the Aichelburg-Sexl metric does not receive
any $\al$ corrections \cite{ama,hs}.
We will argue that this is also true for the
gravitational shock wave in AdS.

To find the CFT dual of this gravitational shock wave, we start in
section 2 by giving the field theory description
of the \w AdS black hole. Since only the AdS metric
is excited on the supergravity side, the only nonzero expectation value
in the CFT is the energy-momentum tensor. It turns out that
the symmetries uniquely determine $<T_{\mu\nu}>$. This is straightforward 
when the CFT is defined on a sphere (cross time), but to describe the boosted
black hole, it is more convenient to work on Minkowski space. So we find
the form of $<T_{\mu\nu}>$ in this case also.

In section 3, we perform the boost and construct the gravitational
 shock wave in AdS.
Using the fact that on the CFT side the boosting corresponds to a Minkowski
dilation,
we find the expectation value of $T_{\mu\nu}$ in
the CFT states which are associated with
the gravitational shock wave. The result is an energy-momentum tensor
which takes the form of null dust  confined to the light cone. We will refer
to these states as ``light cone states".

In Section 4 we use the correspondence between the gravitational shock wave
and these light cone states  to show how the CFT reproduces some basic causal
relations in AdS. In particular, we resolve the two puzzles mentioned above.
The argument that the gravitational shock wave does
not receive $\al$ corrections is given in section 5. Section 6 contains
a discussion of some extensions of our
results, including the possibility of describing the formation of a black hole
from the collision of two null particles in terms of the CFT.

The gravitational shock wave solutions in $AdS_d$ that we construct in
 section 3 might have other implications (which will not be explored here).
For example, it may be used to calculate the amplitude for graviton exchange
 between two massless particles in AdS (in Minkowski space this was done
in \cite{th}) which is essential for the four point  functions \cite{free}. 
Also the shock wave solution in $AdS_7$ can be dimensionally reduced to 
construct the type IIA solution of D0-branes which are localized on D4-branes
in the longitudinal and radial directions and are smeared along the 
angular directions.
These solutions might be useful to study  the $(0,2)$ theory living on 
M5-branes using the DLCQ approach \cite{abs,ah}.  

\sect{CFT description of black holes}

Although most of our discussion applies equally well to AdS in various 
dimensions, to be specific, we will concentrate on the case of 
 $AdS_5\times S^5$. String theory on this space is believed to be
 described by the ${\cal N} =4$ super Yang-Mills (SYM) theory \cite{juan}.
 Since the metric on $S^5$, and the self dual five form will remain unchanged,
 we will concentrate below on the metric on $AdS_5$.

We wish to find the field theory description of the  \w AdS   black hole
 solution:
\be\label{adss}
ds^2=-\left( 1+\frac{\tilde r^2}{R^2}-\frac{8 G M}{3\pi \tilde r^2}\right) 
d\tilde t^2
 +\left( 1+\frac{\tilde r^2}{R^2}-\frac{8 G M}{3\pi \tilde r^2}\right)^{-1} 
{d\tilde r^2}+
\tilde r^2 d\Omega_3^2,
\ee
where $R=(4\pi g_s N)^{1/4}l_s$ is the AdS radius 
and $G$ is the five dimensional Newton's constant.
For finite $N$, the black hole will Hawking radiate and the exact description 
of the system is beyond our current abilities.
In the large $N$ limit (for fixed $g_s N$), 
loop corrections are suppressed and the black hole  becomes classical.
Since $G\sim R^3/N^2$, to have a finite backreaction 
in  the large $N$ limit, $M$ must be of the order $ N^2$. Of course,
we are used to classical black holes having finite mass. But this mass
is in units where $G =1$.  Since $G$ goes to zero in units of the AdS 
radius in the classical limit, we need $M$ to diverge. This implies that
the corresponding
field theory energies are of order $N^2$.
Note that this is typical   for near-extremal D-branes as well.
That is,  the energy above extremality of 
 a  near-extremal D-branes with finite
horizon radius is proportional to $N^2$
\cite{klts,kle}.

On the field theory side, we would like to characterize the states associated
with classical black holes by the expectation value of certain operators.
Since the solution (\ref{adss}) excites only the AdS metric, 
 we conclude, using the arguments
 of \cite{gkp,witten,fer}, that
the only non-vanishing expectation value is of the energy-momentum 
tensor.
We claim that in this case, the symmetries determine
 completely  the form of this expectation value.
This is the field theory analog of the classical 
statement ``black holes have no hair".

The  AdS \w solution (\ref{adss}) breaks the $SO(4,2)$ isometry of AdS to
$SO(4)\times SO(2)$.
This can be seen explicitly 
by relating the above coordinates to the representation of AdS
as the surface
\be\label{a1}
-T_1^2-T_2^2+X_i^2=-R^2,~~~i=1,...,4,
\ee
embedded in a six dimensional flat space with signature $(2,4)$.
In this representation
\be\label{global}
 T_1=(R^2 +\tilde r^2)^{1/2} \sin(\tilde t/R),~~~ T_2=(R^2 +\tilde r^2)^{1/2} 
 \cos(\tilde t/R),
\ee
and the $X_i$ are related to the spherical coordinates in the usual way.
The $SO(2)$ rotation in the $T_1, T_2$ plane is simply translation in
the $\tilde t$ direction.
The $SO(4)$ isometries are rotations in the $X_i, X_j$ planes.

The SYM background associated with the black hole
should have the same unbroken symmetries. In terms of the SYM theory on
$S^3 \times R$, this implies that the expectation value of $\tilde T_{\mu\nu}$ must
be static and homogeneous.\footnote{We use a tilde for the energy-momentum tensor
on $S^3 \times R$ to distinguish it from the one on Minkowski space below.}
Since $\tilde T_{\mu\nu}$ must also be tracefree,
this uniquely fixes the energy-momentum tensor to be
\be\label{tmn}
<\tilde{T}_{\mu\nu}>= \rho\hat{t}_{\mu}\hat{t}_{\nu} +
\frac13 \rho h_{\mu\nu},
\ee
where $\hat{t}_{\mu}$ is a unit vector in the  time direction,
 $h_{\mu\nu}$ is the metric on $S^3$  with radius $R$,
 and $\rho=M/2\pi^2 R^3$,
is the mass density. This is not surprising for large $GM$, since large
black holes are described by SYM states which are approximately thermal.
However, the symmetries require that (\ref{tmn}) is also valid for small $GM$.
\footnote{Small
black holes are known to be unstable to localizing on the $S^5$ \cite{grla},
but if large $N$ SYM is really equivalent to supergravity, it should 
describe all solutions, even unstable ones. The shock wave we construct
in the next section will be stable.}

To describe boosted black holes, it is convenient to reexpress this
energy-momentum
tensor in terms of the theory on Minkowski space. The answer is not
simply (\ref{tmn}) with $ h_{\mu\nu}$ denoting the flat metric on $R^3$.
That would correspond to the near extremal three-brane
geometry which has translational symmetry. We want to describe
the static, spherically symmetric black hole in terms of the theory on 
$R^{3,1}$. There are two ways to do
this. The first is to rewrite the unbroken $SO(2)\times SO(4)$ symmetries
in terms of conformal symmetries of Minkowski space, and look for an
energy-momentum 
tensor invariant under these symmetries. This can be done by
writing AdS in the form
\be\label{adsft}
ds^2 = {U^2\al^2\over R^2}\left( -dt^2 + dx_i^2\right ) + {R^2 dU^2\over U^2},
\ee
where these coordinates are related to the embedding coordinates (\ref{a1})
via
\be\label{bn}
U=\frac{(X_1-T_1)}{\al},~~~
t=\frac{T_2 R}{U \al},~~~x_i=\frac{X_{i+1} R}{U \al},~~~i=1,2,3.
\ee
The SO(3) subgroup of SO(4) implies that the SYM background is spherically 
symmetric. The remaining three generators of $SO(4)$ and the generator of 
$SO(2)$ are linear combinations of  translations and special conformal
transformations. 

An easier way to proceed is to conformally map Minkowski space into 
$S^3 \times R$ and use this to pull back the energy-momentum tensor (\ref{tmn}).
The required conformal factor can be derived by noticing that the metric
on $S^3 \times R$ is obtained by  rescaling AdS in global coordinates
 by $R^2/\tilde r^2$ and taking the limit $\tilde r
\rightarrow \infty$, while the metric on $R^{3,1}$
is obtained by rescaling AdS in coordinates (\ref{adsft})
by $R^2/U^2\al^2$ and taking $U\rightarrow \infty$. So the two boundary metrics are
related by the conformal factor:
\be\label{a3}
\omega^2 =\lim_{U\rightarrow \infty}
\frac{U^2\al^2}{\tilde r^2}=\frac{4 R^4}{(R^2+v^2)(R^2+u^2)},
\ee
where we have written $\tilde r$ in terms of $(U,t,x^i)$ and set
 $u=t-r$, $v=t+r$, with $r^2=x_1^2+x_2^2+x_3^2$.   
In other words, if we start with the Minkowski metric
\be
ds^2=-du dv +\frac{(v-u)^2}{4} d\Omega_2^2,
\ee
and make the conformal transformation $g_{\mu\nu}= \omega^2 \eta_{\mu\nu}$
(with  $\omega$  given by (\ref{a3})), we obtain 
\be
ds^2=-4R^2 d\tilde{U} d\tilde{V} + R^2 \sin^2(\tilde{V}-\tilde{U}) d\Omega_2^2,
\ee
where 
\be\label{a4}
u=R \tan \tilde{U},~~~v=R\tan \tilde{V}.
\ee
Defining $\tilde{V}=\frac12 (\frac{\tilde t}{R}+ \theta)$ and 
$\tilde{U}=\frac12 (\frac{\tilde t}{R}- \theta)$ gives
\be
ds^2=-d\tilde t^2+ R^2(d\theta^2 +\sin^2\theta d\Omega_2^2),
\ee
which is just the metric on $S^3 \times R$.

We now use the following fact (see, e.g., \cite{wald}):
If an energy-momentum
tensor $T_{\mu\nu}$ is conserved and traceless with respect to 
a metric $g_{\mu\nu}$, then $\tilde T_{\mu\nu}$ is  conserved and traceless
with respect to $\tilde g_{\mu\nu} = \omega^2 g_{\mu\nu}$ provided
\be
T_{\mu\nu}= \omega^2 \tilde{T}_{\mu\nu}.
\ee

Using the conformal transformation (\ref{a3}), the coordinate change
(\ref{a4}), and the known form of $<\tilde T_{\mu\nu}>$ (\ref{tmn}),
we obtain the following expectation value for the energy-momentum 
tensor on $R^{3,1}$
\beq\label{aa5}
&& <T_{uu}>=\frac{8 M}{3\pi^2} \frac{R^5}{(R^2+u^2)^3(R^2+v^2)},\non
&& <T_{vv}>=\frac{8 M}{3\pi^2} \frac{R^5}{(R^2+u^2)(R^2+v^2)^3},\\
&& <T_{uv}>=\frac{4 M}{3\pi ^2} \frac{R^5}{(R^2+u^2)^2(R^2+v^2)^2},\non
&& <T_{ij}>=\frac{8 M}{3\pi^2} 
\frac{R^5 \delta_{ij}}{(R^2+u^2)^2(R^2+v^2)^2},\nonumber
\eeq
where $i$ and $j$ denote the orthonormal components on the two sphere.

Small black holes should approximately follow timelike geodesics. Consider
the family of geodesics 
given (in the metric (\ref{adsft})) by
\be\label{geo}
(t^2 +\D r^2 )U^2\al^2 = R^4.
\ee
The parameter $\D r$ labels different geodesics, and determines the maximum
value of $U$ along the curve. The geodesic which stays at the origin $\tilde r=0$
in the global coordinates (\ref{global}) corresponds to $\D r =R$.
These geodesics are all related by a boost
in the  $T_1, X_1$ plane. It follows from (\ref{bn}) that such a boost
corresponds to a  dilation in Minkowski space. The boost which increases
the maximum value of $U$ by $e^\lambda$, decreases distances in $R^{3,1}$
by $e^{-\lambda}$. The energy-momentum
tensor of the boosted black hole can thus  be obtained by simply rescaling
the dimensionful constants $M$ and $R$. 
Setting $E = M e^\lambda$ and $\D r = R e^{-\lambda}$, we obtain
\beq\label{a5}
&& <T_{uu}>=\frac{8 E}{3\pi^2} \frac{\Delta r^5}{(\Delta r^2+u^2)^3
(\Delta r^2+v^2)},\non
&& <T_{vv}>=\frac{8 E}{3\pi^2} \frac{\D r^5}{(\D r^2+u^2)(\D r^2+v^2)^3},\\
&& <T_{uv}>=\frac{4 E}{3\pi^2} \frac{\D r^5}{(\D r^2+u^2)^2(\D r^2+v^2)^2},\non
&& <T_{ij}>=\frac{8 E}{3\pi^2} 
\frac{\D r^5 \delta_{ij}}{(\D r^2+u^2)^2(\D r^2+v^2)^2}.\nonumber
\eeq
One can verify that $E = \int d^3 x <T_{00}>$, so it is indeed
the field theory
energy. It is related to the mass of the black hole by
\be\label{lll}
E=M e^\lambda = \frac{M R}{\D r} = M \sqrt{g_{00}}|_{U_{max}}.
\ee
\begin{figure}
\begin{picture}(150,220)(0,-10)
\vspace{-5mm}
\hspace{30mm}
\mbox{\epsfxsize=120mm \epsfbox{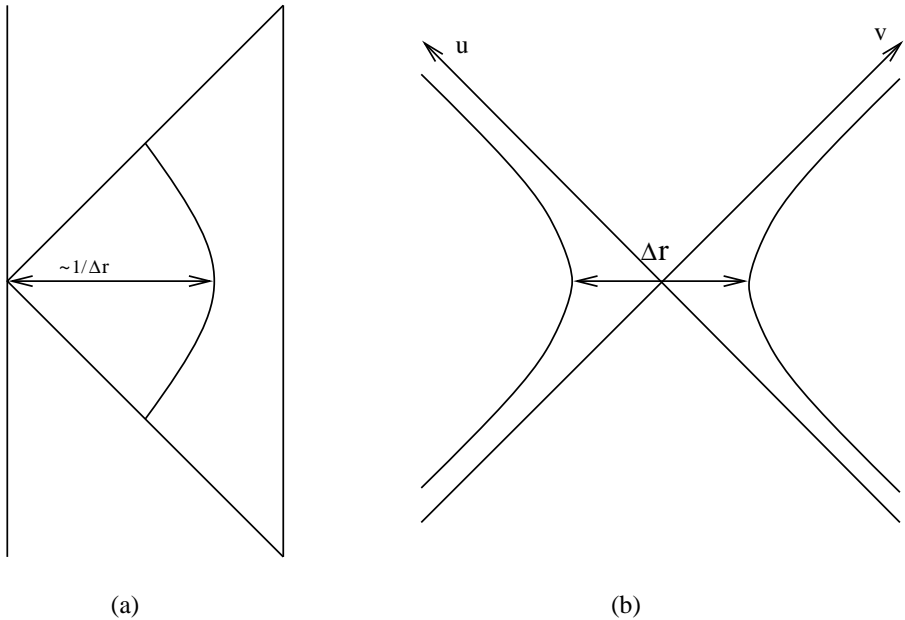}}
\end{picture}
\caption{The UV/IR relation at work:
(a) On the supergravity side a black hole is moving along a radial
 geodesic in AdS with maximal radial direction $\sim 1/\D r$.
(b) On the SYM side, the energy of the SYM state associated with
the black hole is concentrated around the light cone with minimal size of 
$\D r$. }
\end{figure}

Let us describe the profile of the expectation values (\ref{a5}).
The energy-momentum tensor is concentrated near the light cone
$u=0$ or $v=0$ with a width of order $\D r$. So for 
$|t|\gg \D r$ the energy density extends over distances
of the order of $t$ while for $|t|\ll \D r$ the energy
 density is spread over distances of the order $\D r$ (see fig. 1).
This  is in agreement with the UV/IR relation \cite{sus} 
since for $|t|\gg \D r$,
eq. (\ref{geo}) implies that $U\sim R^2 /t\al$ while
for  $|t|\ll \D r$ we get  $U\sim R^2 /\D r\al$ (see fig. 1).

\sect{AdS shock wave solution and light cone states}

Having obtained the SYM description of the Schwarzschild AdS black hole,
we now proceed to construct the analog of the  Aichelburg-Sexl metric
\cite{sexl} describing a gravitational shock wave. On the supergravity
side, this solution can be obtained in two ways. One approach is
to boost the AdS
black hole to the speed of light keeping the total energy fixed.
 Alternatively, one can obtain this solution
using the method described in \cite{dray}, which consists of appropriately
gluing together two pieces of AdS spacetime along a null plane \cite{new2}.
Using these approaches, the AdS shock wave metric was found in
four dimensions in \cite{swads} and further studied in \cite{pogr}.
Below, we reexpress this solution in new coordinates
which are more convenient for our purposes, and generalize it to 
higher dimensions. We then find the  SYM description
of this solution  by applying the Minkowski dilation which is the analog of 
the AdS boost. To simplify the equations,
in this section, we set $R^2=\al=1$. 

To construct the gravitational shock wave we find it useful to work with the 
coordinate system
\be\label{lim}
y_0=\frac{ T_1}{1+T_2},~~~~ y_i=\frac{ X_{i}}{1+T_2}.
\ee
In these coordinates, the $AdS_d$ metric takes the form
\be\label{1}
ds^2_0=\frac{4  \eta_{\mu\nu} dy^\mu dy^\nu}{(1-\eta_{\a\b} y^\a y^\b)^2},
\ee
where $\eta_{\mu\nu}$ is the usual $d$  dimensional Minkowski metric.
The physical spacetime corresponds to the region 
$\eta_{\mu\nu} y^\mu y^\nu < 1$ and the
boundary at infinity is  $\eta_{\mu\nu} y^\mu y^\nu =1$. 
This clearly shows that $AdS_d$
is conformal to the region of $d$ dimensional Minkowski space lying
inside a timelike hyperboloid. (This should not be confused with
the $d-1$ dimensional Minkowski space of the field theory.) These coordinates
do not cover the entire spacetime, but only the shaded region
shown in Fig. 2a.
The advantage of this coordinate system is that  the metric is
manifestly Lorentz invariant, though not static.
This makes the form of the $AdS$ shock wave clearer
 knowing the shock wave solution in Minkowski space.

\begin{figure}
\begin{picture}(150,250)(0,0)
\vspace{-5mm}
\hspace{30mm}
\mbox{\epsfxsize=110mm \epsfbox{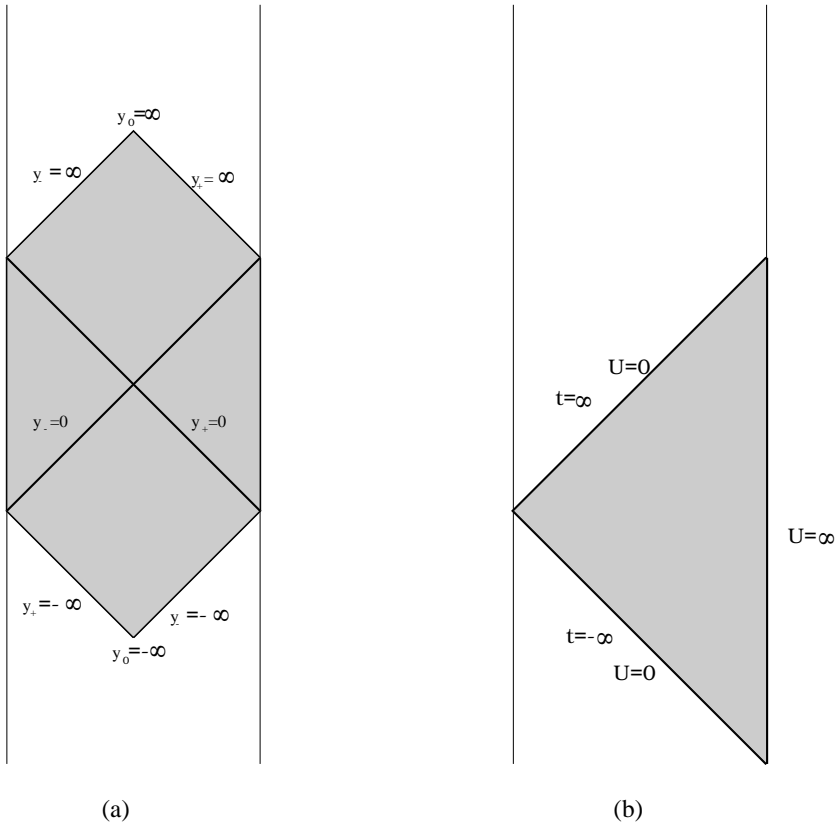}}
\end{picture}
\caption{Regions of $AdS_d$ covered by different coordinates:
The coordinates (\ref{lim}) cover the shaded region in (a).
While the coordinates (\ref{bn}) cover the shaded region in (b).
The matching between the coordinates is such that the line $U=0, t=\infty$
in (b) is the 
same as $y_-=0$ in (a). }
 \end{figure}

We would like to find the metric in the presence of a massless particle which
 moves along the null geodesic $y_0 +y_1=0$.
By analogy with the solution for zero cosmological constant,
we try a metric of the form
\be\label{b4}
 ds^2 = ds^2_0 + \frac{p\ \delta(y_+)
 f(\rho) dy_+^2 }{ (1+y_+ y_- - \rho^2)},
\ee
where
\be
y_-=y_0-y_1,~~~y_+=y_0+y_1,~~~\rho^2=\sum_{i=2}^{d-1}y_i^2.
\ee 
The reason for the  factor $1+ y_+ y_- - \rho^2$ in the dominator is
 that it simplifies the equation for $f$. 
 Substituting (\ref{b4}) 
into Einstein's equation 
$R_{\mu \nu}=-(d-1)g_{\mu \nu}$ yields the following linear equation for $f$:
\be\label{linear} 
D^2 f - 4(d-2) f =0,
\ee
where $D^2$ is the Laplacian on the transverse surface of constant $y_-$
and $y_+=0$ which is  just $d-2$ dimensional hyperbolic space
with metric
\be
ds^2 = {d\rho^2 + \rho^2 d\Omega^2 \over (1-\rho^2)^2}.
\ee
To describe the field of a massless particle in AdS, one should add a
delta function source to the right hand side of (\ref{linear}).  Near
$\rho =0$ the cosmological constant is negligible and the
solution will resemble the Aichelburg-Sexl metric.

The fact that $f$ satisfies a linear equation  implies that,
 like in Minkowski space, two parallel massless particles do not interact.
That is, a massless particle moving along the
geodesics $y_+=0, y_i=0,$ for $i\neq 1$ will not interact with
a massless particle moving along, say, $ y_+=0,$ $ y_2=\mbox{const.}, y_i=0
$ for $i\neq 1, 2$.

When the cosmological constant is zero, the gravitational shock wave solution
preserves half the supersymmetry \cite{tod}. We believe that the same is
true for negative cosmological constant, although we have not yet
checked this. The ability to superpose solutions is  strong evidence
for this.

We now turn to the SYM description of this gravitational shock wave.
To relate string theory in $AdS$ to SYM
on $R^{d-1}$ it is convenient to use the $(U,t,x^i)$ coordinates
(\ref{bn}).
The relation between these coordinates and $y^\mu$ (\ref{lim}) is 
\be\label{uy}
U=-\frac{2y_-}{1+y_-y_+-\rho^2},~~~t=-\frac{1-y_-y_++\rho^2}{2y_-},~~~x_i=
-\frac{y_{i+1}}{y_-}.
\ee
So the null geodesic $y_+=0$, $\rho =0$ is given by
\be\label{c3}
U=1/t , ~~~x_i=0,~~~\forall i.
\ee
We see that  at $t=0$ the particle is located at the boundary and it falls
 towards $U=0$ at $t=\infty$.  

In the previous section we found the SYM description of a particle (small
black hole) which
 follows the geodesic (\ref{geo}). 
The geodesic (\ref{c3}) is obtained from (\ref{geo}) by taking the limit
$\D r\r 0$. (Recall that we have set $R^2 =\al$.)
We want to take this limit keeping the energy fixed.
Therefore, the field theory energy-momentum tensor associated with the null 
particle is obtained from (\ref{a5}) by taking the limit $\D r\r 0$ 
keeping  $E$ fixed. It is easy to check that the
total field theory energy, $E=\int d^3 x <T_{00}>$,
does not depend on $\D r$, so the limit is straightforward.
 It is clear that for $u,v \ne 0$,
 $\lim_{\D r\r 0}<T_{\mu\nu}>=0$. So the energy-momentum tensor is a 
 delta-function supported
 the light cone, $u=0$ or $v=0$. Furthermore, on the future light
 cone, $u=0$, the only nonzero component is $<T_{uu}>$. 
Similarly, on the past light cone, $v=0$, the only nonzero component is
$<T_{vv}>$. This implies that the energy-momentum tensor takes the form
of null dust: $<T_{\mu\nu}> \propto l_\mu l_\nu$ where $l^\mu$ is
a null vector tangent to the light cone. 
Since $<T_{\mu\nu}>$ is 
nonzero only on the light cone, we will refer to these states
as ``light cone states".

We now review a field theory argument  due to Coleman and Smarr \cite{coleman}
that supports our result that the energy-momentum tensor is localized on the light
cone.
Consider any classical field theory whose energy-momentum tensor satisfies
\beq\label{12}
&& T_{00}\geq 0,\non
&& \partial _{\nu} T^{\mu\nu}=0,\\
&& T^{\mu}_{\mu}=0 \nonumber.
\eeq
  From the moment of inertia $I(t)=\int d^3x\  r^2\ T_{00}$,
one can define the average size of a state  at a given
 time by 
\be\label{13}
\bar{r}(t) ^2=I(t)/E, ~~~~\mbox{where}~~~
E=\int d^3x \ T_{00}.
\ee
Using eqs. (\ref{12},\ref{13}) one finds that $d^2  \bar{r}^2/ dt^2=2 $
which implies that 
\be\label{14}
\bar{r}^2=t^2+\D r^2,
\ee
where $\D r$  is the minimum average size (which we assume occurs at $t=0$).
This means that the initial configuration expands very rapidly, with its average
size always larger than a sphere of light emitted at the origin at $t=0$.
In the limit $\D r \r 0$, the initial configuration becomes localized at 
a point. Causality then requires that the fields  vanish outside
the light cone. Since $\bar{r}^2=t^2$, the energy density must be localized on
the light cone.

The above argument was intended for a classical field theory. However the
second two conditions of (\ref{12}) are satisfied by the expectation
value of the energy-momentum tensor in any conformal field theory.
Furthermore, even though the first condition may be violated, as long
as the total energy remains positive, we can still apply the argument.
This is because a negative energy density will only decrease the average size
$\bar r$, but the conclusion that $\bar r\ge t $ will still hold.
Thus one can also apply this argument in the quantum theory. Even though
the state is localized to a point at $t=0$, there is no contradiction 
with the uncertainty principle since it has essentially infinite energy: As
discussed in the beginning of section 2, keeping 
a  finite backreaction on the supergravity side in the classical limit
$N\r\infty$ requires that
$E\sim N^2$.

These light cone
states are probably also supersymmetric. One usually thinks that
supersymmetric states must be static, since the square of the supersymmetry
generator is usually a time translation. 
The light cone states are clearly not static, but in a superconformal
theory, one has the possibility that the supersymmetry generator
will square to a special conformal transformation (see, e.g., \cite{min}).
 One can
verify that the light cone states are 
invariant under the special conformal transformations  
by translating the symmetries of the  null geodesic (\ref{c3}) to
the field theory.

We have been describing these states in terms of the field theory on Minkowski 
space. If one conformally rescales to $S^3 \times R$, the light cone states start
 at a point, expand along the light cone to a maximal size at the equator, and
 then collapse back to a point at the other side of the sphere, where the
 particle again hits the boundary.
Only the first half corresponds to the $R^{(3,1)}$ description.
The second half can be described by a time reversed process on a second 
copy of the Minkowski space.

\sect{Tests of causality}

We now apply the correspondence between the gravitational shock waves and
the  light cone states to test whether SYM reproduces the causal 
relations in $AdS$. To begin,
 consider two massless particles in $AdS$ which come in from
infinity from the same direction, but at different times. In other words,
they follow the null geodesics $y_+=0, \rho=0$ and $y_+=a>0, \rho=0$.
The first particle produces a gravitational field described by the
shock wave metric (\ref{b4}). 
The second particle produces an analogous shock wave
solution obtained from (\ref{b4})
by the coordinate change $y_+\r y_+ -a$.
Since the nontrivial curvature in the first shock wave is concentrated
 on the null plane $y_+=0$, and in the second is concentrated on $y_+=a$,
 we can clearly combine them into a single solution describing both
 particles. Since the shock waves do not intersect, there is no interaction
 between these particles.
 In the SYM theory, we have seen that the corresponding
states are localized on the light cone, so once again, they do not
interact (see fig. 3). 

Note  that the absence of interactions is a result of having 
 massless particles. Suppose we consider,
 following \cite{bdhm}, two strings which wind
around the $x_3$ direction and come in from infinity from the same point (in
the  $x_1$ and $x_2$ directions) at different times.
On the supergravity side, the backreaction of such  strings is not confined to 
 null surfaces and 
hence they will interact.
On the field theory side, the wound strings are represented by 
flux tubes which expand in time.
The expansion in this case is not confined to the light cone and hence the
flux tubes will also interact.

\begin{figure}
\begin{picture}(100,220)(60,0)
\vspace{-5mm}
\hspace{30mm}
\mbox{\epsfxsize=40mm \epsfbox{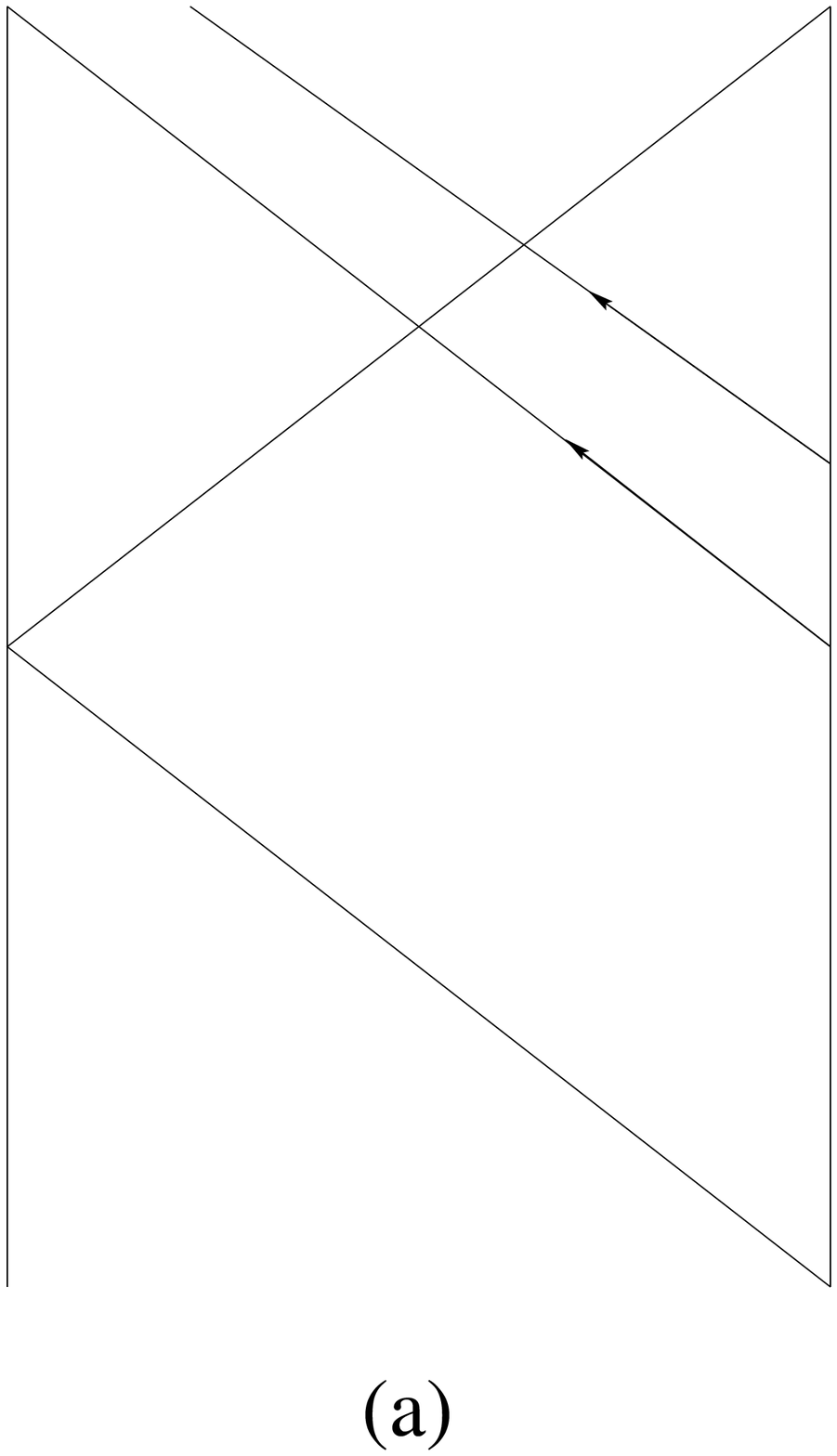}}
\end{picture}
\begin{picture}(100,180)(-30,0)
\vspace{-5mm}
\hspace{30mm}
\mbox{\epsfxsize=60mm \epsfbox{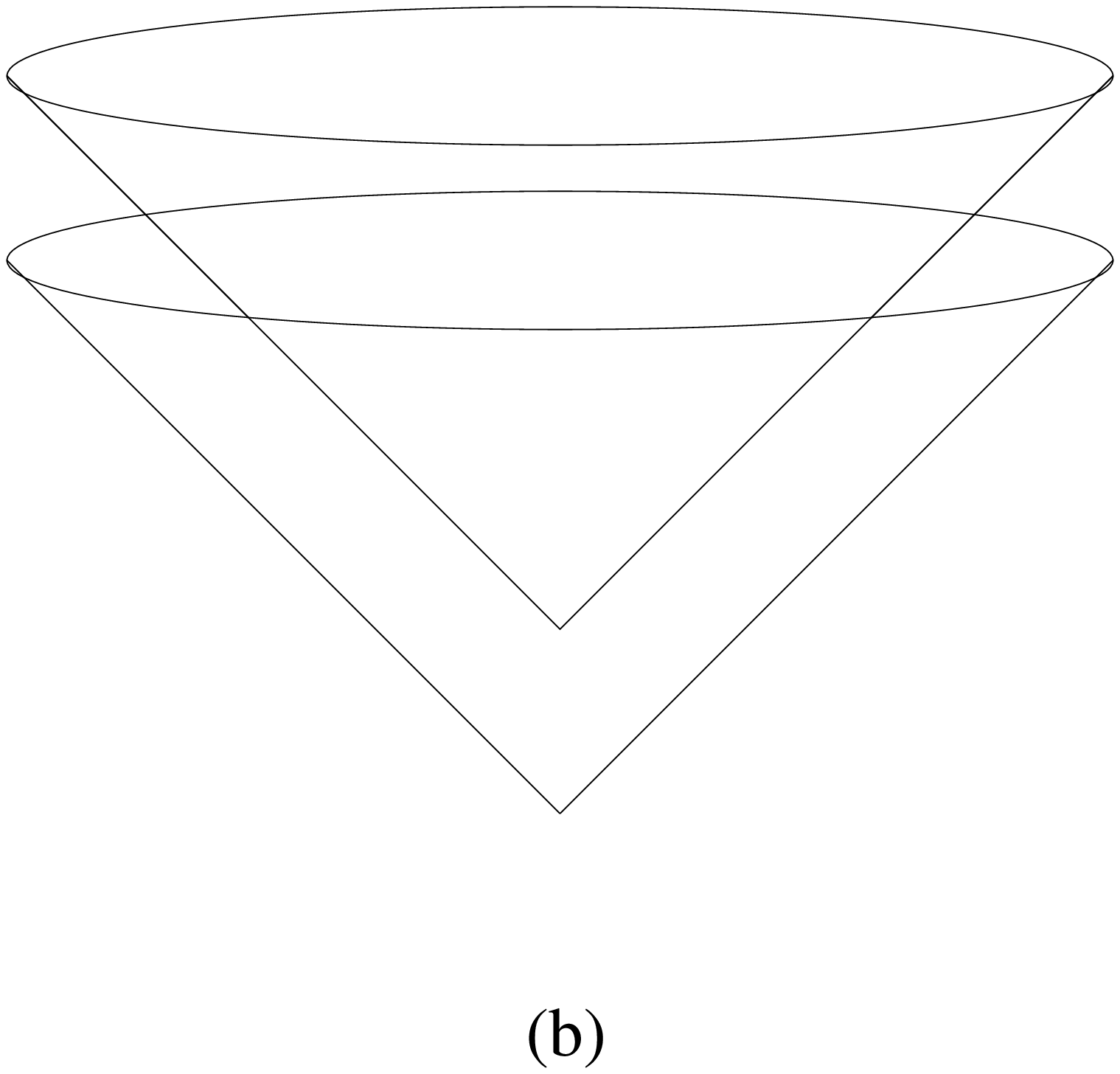}}
\end{picture}
\caption{On the supergravity side (a) we have two shock waves  which
 do not interact. On the SYM side (b) the states associated with these 
 shock waves evolve on light cones which do not cross. }
\end{figure}

Another case in which we know that there are no interactions on the
supergravity side is of parallel massless particles.
To be specific we consider two massless particles following the geodesics
\beq \label{paral}
I) && ~~~y_+=0,~~~ y_i=0 ~~\forall i, \non 
II) && ~~~y_+=0,~~~ y_2=\Lambda,~~~y_i=0 ~~~\mbox{for}~~ i\neq 2.
\eeq
In the $(U,t,x^i)$ coordinates, these worldlines are described by
\beq
I) && ~~~ t=1/U,~~~x_i=0, \non
II) && ~~~ t=A/U,~~~x_1=B/U, ~~~x_i=0 ~~\mbox{for}~~ i>1,
\eeq
where $A=\frac{1+\Lambda^2}{1-\Lambda^2}$ and $B=\frac{2\Lambda}{1-\Lambda^2}$.
Notice that at the boundary ($U=\infty$), $t_{I}=0 = t_{II} $ and 
$x_{iI}=0=x_{iII}$.
In other words, even though the proper distance between the two particles
is constant in $AdS$ (as follows from (\ref{1})) their separation
in $(t,x^i)$ goes to zero since $U^2$ blows up at the boundary.
Thus from the SYM point of view, these two particles are described by
states  localized on the same light cone. It is certainly not obvious why they
do not interact. Although we do not have a complete answer to this puzzle,
this result can be made plausible as follows.
Since $A^2-B^2=1$, the SYM  states associated with the two particles are related
 to each other via 
a boost in the $(t, x_1)$ plane.
(Note that the conclusion of the previous section that the states evolve on
 the light cone is invariant under the boost.)
Classically, two massless particles starting from the same point but in
 different directions will not interact since 
they are not causally connected.
The fact that the energy-momentum tensor of the light cone states takes the
form of 
null dust strongly supports the idea that these states will not interact.
If the light cone states are indeed supersymmetric and analogous to BPS
states, this would provide another argument for the absence of interactions.

Next we consider a typical case in which there are interactions.
We wish to compare the minimum time for interactions to occur
in both the SYM and supergravity description.
Consider two massless particles which come in from infinity
at the same time, but from different directions in space.
They are described by the null geodesics
\beq\label{21}
&& I) ~~~~~ U=1/t,~~x_i=0 ~~ \forall i, \non 
&& II) ~~~~~ U=1/t,~~x_1=c, ~~x_i=0~~ \mbox{for}~~ i>1.
\eeq
In terms of SYM we have two states which are evolving on the  light cones
\beq\label{lcs}
&& I) ~~~~~ t^2-\sum_{i}x_i^2=0\non
&& II) ~~~~~ t^2-(x_1-c)^2-\sum_{i>1}x_i^2=0.
\eeq
Clearly, these states will start to interact at a  time $t=c/2$ when
the light cones intersect (see fig. 3b). This seems to contradict the
fact that the particles in (\ref{21}) do not intersect for any
finite $t$. (The particles do intersect at $U=0$, but this corresponds to 
infinite $t$.) As we have discussed, the
resolution  is that the SYM state describes not
 only the motion of the particle 
but its gravitational back-reaction (that is, the  shock wave) as well.
We need to find, therefore, when the shock waves cross each other.
\begin{figure}
\begin{picture}(100,220)(60,0)
\vspace{-5mm}
\hspace{30mm}
\mbox{\epsfxsize=50mm \epsfbox{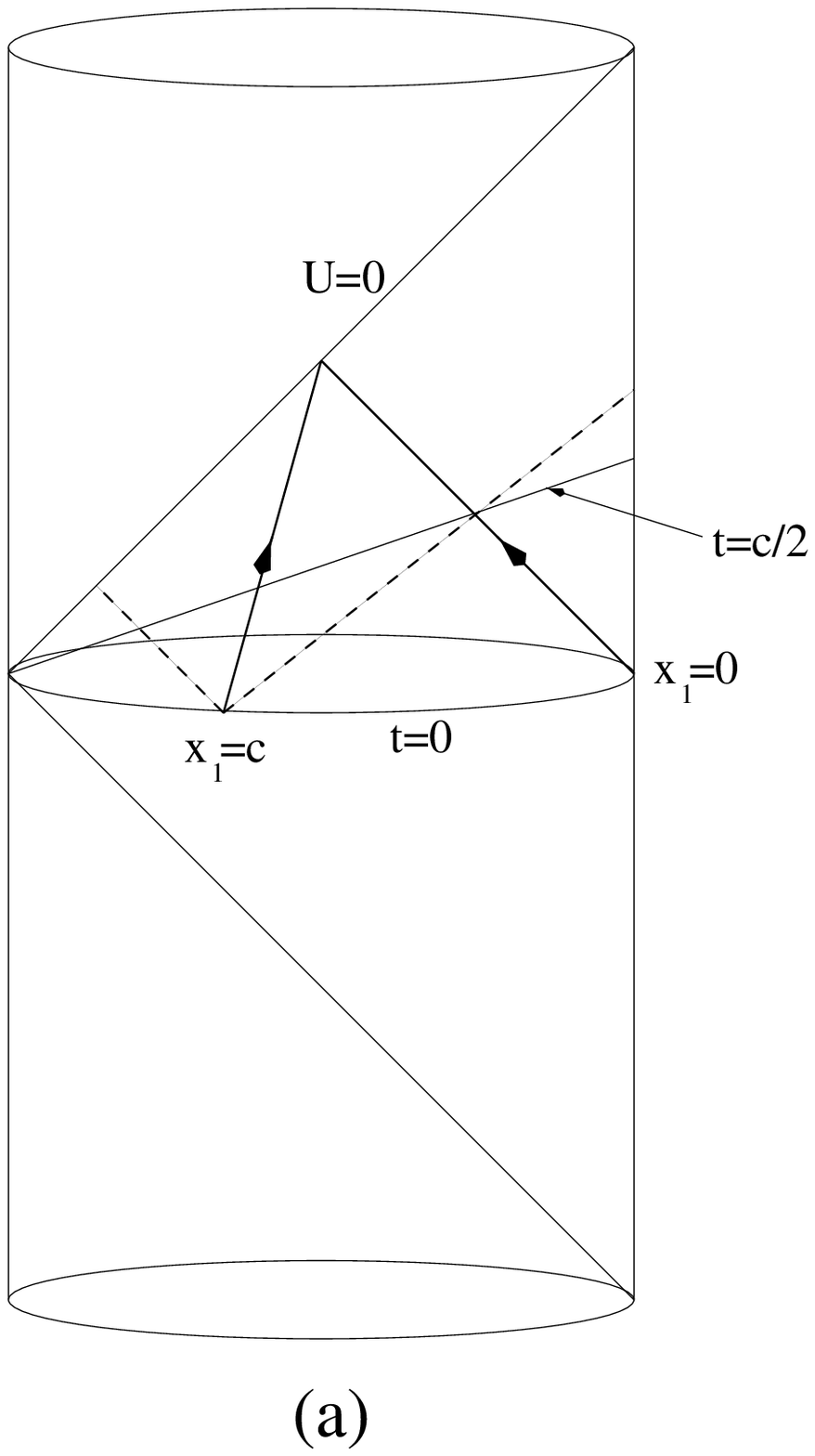}}
\end{picture}
\begin{picture}(100,180)(-30,0)
\vspace{-5mm}
\hspace{30mm}
\mbox{\epsfxsize=70mm \epsfbox{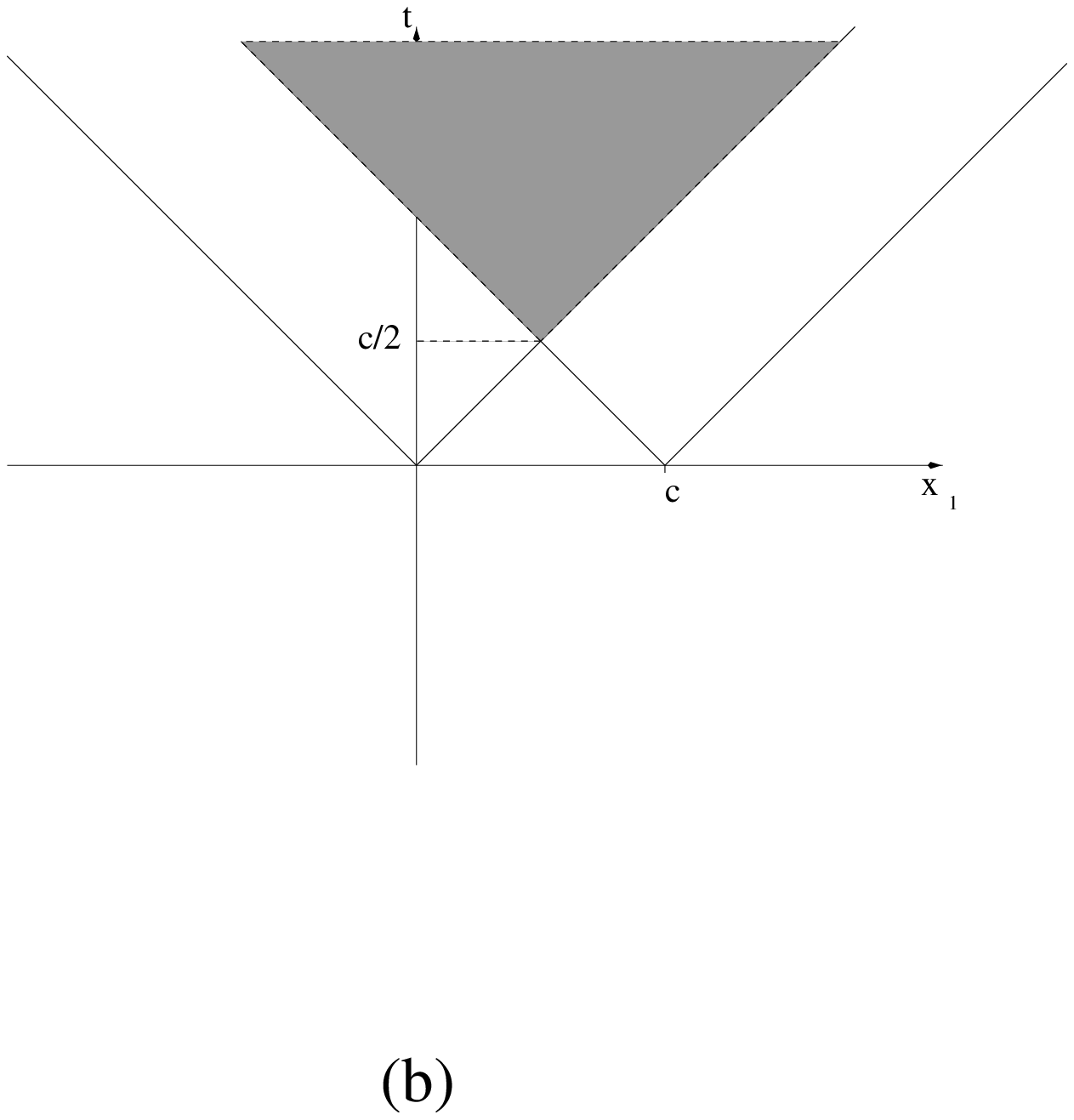}}
\end{picture}
\caption{ (a) The arrows indicate the trajectories of the  massless particles.
 The dashed lines indicate the shock wave of the particle whose trajectory
initiates at $x_1 =c$. 
The shock wave of the other particle coincides (in the figure) with its
 trajectory. The shock waves cross each other at $t=c/2$ which is in 
agreement with  SYM (b).  }
\end{figure}

The exact solution for two intersecting shock waves is not known 
completely\footnote{For a discussion of the analogous solution with
zero cosmological constant, see \cite{depa}.}, 
but fortunately, it is
not needed to answer this question. The complicated part of the solution
lies to the future of the intersection of the shock waves. Elsewhere,
the solution is simply AdS with two approaching waves. We want
to compute the time (in the $(U,t,x^i)$ coordinates associated with the
Minkowski time in the field theory) for the shock waves to meet.
However it is easier to describe the approaching
shock waves in $y^\mu$ coordinates, since they lie along planes in these
coordinates. So we will first use the $y^\mu$ coordinates, and then
translate back using (\ref{uy}). We have seen that
the solution for the gravitational field induced by particle (I) in (\ref{21})
has a shock along the plane $y_0+y_1=0$. The region that overlaps the
$(U,t,x^i)$ coordinates with $U>0$ has $-\infty <y_0 < 0$.
Particle (II) follows the
curve $y_+ = c^2 y_-, \ y_2 = -c y_-, \ y_i =0$ for $i\ne 1,2$.
This implies $y_2 =2 c y_1/(1-c^2)$ so this worldline can be obtained from
the original one (satisfying $y_2=0$) by
 a rotation in the $y_1,y_2$ plane.
So the plane of its shock wave will be similarly rotated:
$y_0+\cos\alpha\ y_1+\sin\alpha\ y_2=0$ where $\cos\alpha =
\frac{1-c^2}{1+c^2}$.
Therefore,
the two shock waves  will cross each-other and interact at 
\be\label{inter1}
-\infty \leq y_0 \leq 0,~~~y_1=-y_0, ~~~y_2=-cy_0, 
\ee
and hence the field theory time of intersection is 
\be\label{inter2}
t\geq - \frac{1+y_0^2 c^2}{4y_0}\geq c/2,
\ee
which is in a precise agreement  with the SYM results. 

This agreement actually reflects a deep connection between the causal
structure of $AdS$ and its boundary.
The conformally completed $AdS$ spacetime resembles
a solid cylinder. But there is a crucial difference between the
causal structure of this spacetime and a solid (timelike) cylinder in
Minkowski spacetime. The intersection of a  null plane in Minkowski spacetime
and a timelike cylinder is a spacelike curve. Similarly, a null curve
which stays on the boundary of the cylinder is not a null geodesic, and
reaches the other side at a
later time than a null geodesic which passes through the interior. In $AdS$
this is not the case. A null geodesic which starts on the boundary
stays on the boundary. A null plane in the interior intersects
the boundary at infinity in a null surface which is, in fact, the
future light cone\footnote{We could have said `past light cone'. Since
all null geodesics from a point $p$ on the boundary
focus on the other side of the $S^3$ at the same point $q$, the future 
light cone of $p$ is the same as the past light cone of $q$.} 
of a point $p$. In fact, the null plane is simply part of the
future light cone of $p$ consisting of those null geodesics which enter the 
interior. 

This explains the agreement we just found.  The nontrivial curvature
in the gravitational shock wave lies on a null plane which is part of
the future
light cone of a point on the boundary at infinity. So the intersection of two
shock waves occurs first where their light cones intersect, which is 
on the boundary. Indeed, the time $t=c/2$ corresponds to $y_0=-1/c$ in
(\ref{inter2}) which implies $U=\infty$ using (\ref{inter1}) and (\ref{uy}).
Hence the time is the same as if one first restricts the
light cones to the boundary and then asks when they intersect (which is
just the field theory answer). Note that  this argument is about
the relation between the causal structure in the interior and on
the boundary. The nontrivial curvature in the
shock wave metric goes to zero for large distance from the source,
so there is no `extra curvature' on the boundary light cone.
However, for any $t>c/2$ the shock waves will intersect in the interior
resulting in gravitational interactions.

\sect{$\al$ corrections}

In this section we show that there are no $\al $ corrections to the AdS shock
 wave solution.
The fact that the solution probably breaks 1/2 of the supersymmetries does
not necessarily imply that there are no $\al$ corrections,
since usually higher order corrections do modify BPS solutions
(though they do not modify the relation between the charges and the energy).
However, there are several examples when higher order terms do not modify
the solution at all. Our argument for the AdS shock wave
solution will be based on two such examples. The first is the
$AdS_5 \times S^5$  metric which has recently been shown to be an
exact solution \cite{kr}. The second is the
shock wave solution in Minkowski space.\footnote{
Another example, which is not related to our discussion, is of the SYM 
solution associated with of an electric charge, which is realized
in string theory as a string attached to the D-brane.
In  \cite{cal} it was shown that the DBI corrections 
to SYM  do not modify the solution.}
In \cite{ama} it was shown that this background
is a solution of the $\sigma$-model equation of motion to all orders
in perturbation theory.
A geometrical derivation of this result, which rests on the fact that the 
curvature is null, was given in \cite{hs}.
The result of \cite{ama} cannot be generalized to AdS since we do not know
 how to describe string theory with a RR-background  by a $\sigma$-model.
However, the geometrical approach of \cite{hs} can be generalized to AdS 
to show, together with the result of
\cite{kr}, that higher order (local) corrections will not modify the solution.

To see that there are no $\al$ corrections, let us write the
 AdS shock wave solution (\ref{b4}) in the form
 \beq
 g_{\mu\nu} =  g_{\mu\nu}^0 + F l_\mu l_\nu
 \eeq
 where $g_{\mu\nu}^0$ is the AdS metric and
 $l_\mu=\partial_\mu y_+$. Then $l^\mu l_\mu =0$ and the
Riemann curvature takes the form
\beq
&& R_{\mu\nu\lambda\sigma}=R_{\mu\nu\lambda\sigma}^0+
R_{\mu\nu\lambda\sigma}^1,\non
&& R_{\mu\nu\lambda\sigma}^0=-(g_{\mu\lambda}g_{\nu\sigma}-g_{\mu\sigma}
g_{\lambda\nu}),\\
&& R_{\mu\nu\lambda\sigma}^1=l_{[\mu}K_{\nu] [\lambda}
 l_{\sigma ]},\nonumber
\eeq
$R_{\mu\nu\lambda\sigma}^0$ is the usual AdS curvature, and $K_{\mu\nu}$
is a symmetric tensor satisfying $K_{\mu\nu} l^\nu =0$. 
The  $\al$ corrections
are derived from terms in the action  involving higher powers and derivatives
of the curvature. Consider first a scalar constructed just from powers of the
curvature. It is clear that $R_{\mu\nu\lambda\sigma}^1$ cannot contribute
since the null vectors must contract either on the metric, on  $K_{\mu\nu}$,
or on themselves, and in all cases, the result is zero.
Terms involving covariant derivatives of the curvature
will also not contribute since the
covariant derivative of $R_{\mu\nu\lambda\sigma}^0$ is zero and one can show
that $\nabla _\mu l_\nu = V_{(\mu} l_{\nu)}$ for some vector $ V_\mu$
which is orthogonal to $l^\mu$.
Thus one cannot get rid of the factors of $l^\mu$ in $R_{\mu\nu\lambda\sigma}^1$
by taking its covariant derivative. Since covariant derivatives
of $K_{\mu\nu}$ are also orthogonal to $l^\mu$,  all contractions
of $l^\mu$ will again vanish. Thus the only possible
$\al$ correction terms come from $R_{\mu\nu\lambda\sigma}^0$.
But these should be identical to the $\al$ corrections of AdS itself
which have been shown to vanish \cite{kr}.
We conclude, therefore, that there are no $\al$ corrections to the AdS
shock wave solution.
   
\sect{Discussion}

We have found the description of a static spherically symmetric black hole
in AdS in terms of the SYM theory on Minkowski space. By boosting the
black hole to the speed of light keeping the total energy fixed, one
obtains a gravitational shock wave in AdS. We have seen that 
the analogous procedure on the
SYM side yields light cone states -- states whose energy-momentum tensor
is localized on the light cone. Using the duality between gravitational
shock waves and light cone states, we have shown how the SYM reproduces
some basic causal properties of AdS.

There are many open questions. In addition to the obvious one of testing
other aspects of causality in AdS, one can ask if this analysis extends
to the nonconformal dualities associated with other D-branes \cite{imsy}. 
It should
be possible to introduce gravitational shock waves in any supergravity
background. Are there analogs of the light cone states in the corresponding
field theories?

One can imagine other applications of the duality between shock waves
in AdS and light cone states.
For example, the collision of two gravitational
shock waves should produce a black hole. Since we know the SYM
description of both the shock waves and the black hole, can one
see this process in the SYM theory? In the example given in (\ref{21}),
the particles did not collide until an infinite field theory time,
so one could never see the formation of a black hole in this case. 
However, one can consider other null geodesics which collide 
at finite time, such as
\beq
&& I)~~~~~t=\sqrt{1+\lambda^2}/U,~~x_1=\lambda/U,~~~x_i=0~~\mbox{for}~~~ i>1,\non
&& II)~~~~~t=\sqrt{1+\lambda^2}/U,~~x_1=-\lambda/U+c,~~~x_i=0~~\mbox{for}~~~ i>1,
\eeq
 where $c>0, \lambda >0$. (The case $\lambda=0$ corresponds to (\ref{21}).)
These geodesics intersect in a time
\be
t=c\frac{\sqrt{1+\lambda^2}}{2\lambda}.
\ee
 
On the SYM side the two particles are described by two states evolving 
on the light cones (\ref{lcs}). These light cones
always intersect after a time $c/2$ independent of  $\lambda$. So the 
formation of a black hole is not directly related to this intersection.
It seems to depend on concentrating the energy in the SYM.
For $\lambda =0$ the energy density on each light cones is spherically
symmetric before they intersect.
The effect of  $\lambda > 0$ is to boost the light cone states toward each
other, and hence concentrate the energy in the intersection region.

One might wonder how the energy-momentum tensor could ever evolve into
the black hole form (\ref{a5}) which is nonzero everywhere, when causality
requires that it be zero outside the future of the two light cones (\ref{lcs}).
The answer is simply that the
collision of shock waves in AdS will take a long
time to settle down to a static black hole. It is easy to see that this
will not occur in a finite field theory time $t$. For large $\lambda$,
one should be able to see the energy-momentum tensor start to approach
the form (\ref{a5}) at late times. 

It may be possible to make more progress on this question by considering
the special case of $AdS_3$. One unusual feature of this case is that
since the field equations require the solution to be locally $AdS_3$
away from the source, there is no gravitational shock wave solution.
The solution describing a null particle in $AdS_3$ just has a 
conical singularity along a null geodesic \cite{mats}.
It is not clear if this should
be described by a 2d analog of the light cone states, which would have
support on two null curves. One advantage of the low dimensions is that
it is possible to find
an exact supergravity solution describing the formation of a black hole
from the collision of two null particles \cite{mats}.
It would be interesting to 
find the CFT description of this process.

\vspace{1cm}
\centerline{\bf Note added}
After this paper was written we learned of \cite{new1} which has some overlap
 with section 2.

\vspace{1cm}
\centerline{\bf Acknowledgments}

We would like to thank A. Hashimoto, V. Hubeny, J. Polchinski and E. Witten
for discussions.
This work was supported in part by NSF grants PHY95-07065 and
PHY97-22022.

  \end{document}